\documentclass[fp,twocolumn]{jpsj3}
\usepackage{txfonts}
\usepackage{graphicx,bm,color,braket}

\title{Vortex-Core Charging Due to the Lorentz Force in a $d$-Wave Superconductor}

\author{Hikaru Ueki, Wataru Kohno, and Takafumi Kita}
\inst{Department of Physics, Hokkaido University, Sapporo 060-0810, Japan} 

\abst{
We derive augmented quasiclassical equations of superconductivity with the Lorentz force in the Matsubara formalism 
so that the charge redistribution due to supercurrent can be calculated quantitatively.
Using it, we obtain an analytic expression for the vortex-core charge of an isolated vortex in extreme type-II materials 
given in terms of the London penetration depth and the equilibrium Hall coefficient.
It depends strongly on the Fermi surface curvature and gap anisotropy, and may change sign 
even as a function of temperature due to the variation in the excitation curvature under the growing energy gap.
This is also confirmed in our numerical study of high-$T_{\rm c}$ superconductors.
}

\begin{document}
\maketitle

\section{Introduction \label{sec:I}}
It was pointed out  two decades ago that vortex cores in type-II superconductors, 
each of which embraces a single magnetic flux quantum, may also accumulate charge  \cite{KF,Feigel'man}.
Since then, extensive studies have been carried out both theoretically \cite{Blatter,Hayashi,ESR,MH,Chen,Knapp} and experimentally \cite{Kumagai},
especially in connection with the sign change of the flux-flow Hall conductivity observed in 
a number of type-II superconductors \cite{GZ,Iye,AGL,Hagen1,Hagen2,Chien,Luo}.
The purpose of this work is to develop a theoretical formalism to investigate the charging microscopically in detail and 
present a calculation of the charge redistribution around an isolated $d$-wave vortex with anisotropic Fermi surfaces.

The topic of charging in superconductors may be traced back to the pioneering work by London \cite{London,Kita09} 
when he included the Lorentz force in his phenomenological equations of superconductivity. 
He thereby predicted the emergence of net charging due to the Hall effect whenever supercurrent flows. 
On the other hand, early studies on vortex-core charging \cite{KF,Feigel'man,Blatter} 
regard the core as a normal region and consider its chemical potential difference from the surroundings due to the particle-hole asymmetry in the density of states.
Thus, one may wonder how these two apparently different approaches may be connected with each other microscopically.
Although the core charging itself has been confirmed by more refined calculations based on the Bogoliubov--de Gennes equations \cite{Hayashi,MH,Chen,Knapp}, 
it would be useful to have a formalism that enables us to calculate the charging easily
for anisotropic Fermi surfaces and/or energy gaps continuously from the outer region into the core center.
Note in this context that the Fermi surface curvature is a crucial element for determining the sign of the normal Hall coefficient.

Suitable to this end may be the quasiclassical Eilenberger equations \cite{Eilenberger}.
Indeed, they have been used extensively to study vortices quantitatively \cite{KP,Klein,SM,Ichioka1}
and are now regarded as a basic  and reliable tool for investigating inhomogeneous and/or nonequilibrium superconductors microscopically \cite{SR,LO86,Kopnin,KitaText}.
However, the standard equations cannot describe the charging because of the missing Lorentz force, 
which has been incorporated successfully in a gauge-invariant manner within the real-time Keldysh formalism \cite{Kita01}.
The augmented quasiclassical equations in the Keldysh formalism 
have been used to study charging in the Meissner state with Fermi surface and gap anisotropies \cite{Kita09},
and also to calculate flux-flow Hall conductivity numerically for the $s$-wave pairing on an isotropic Fermi surface \cite{AK}. 
On the other hand, it is still desirable when studying the charging to transform the equations into the Matsubara formalism,
in which equilibrium properties and linear responses  can be calculated much more easily.
We carry this out below so that microscopic and quantitative calculations of charging in inhomogeneous superconductors become possible,
as exemplified in our model $d$-wave calculations presented below.

This paper is organized as follows. 
In sect. \ref{sec:II}, we derive the augmented quasiclassical equations of superconductivity with the Lorentz force in the Matsubara formalism. 
In sect. \ref{sec:III}, we study the analytic continuation between the Matsubara and Keldysh Green's functions
obeying the augmented quasiclassical equations. 
In sect. \ref{sec:IV}, we derive an analytic expression for the vortex-core charge. 
In sect. \ref{sec:V}, we present numerical results for vortex-core charging.
In sect. \ref{sec:VI}, we provide a brief summary.

\section{Augmented Quasiclassical Equations in the Matsubara Formalism \label{sec:II}}

First, we derive the augmented quasiclassical equations of superconductivity with the Lorentz force in the equilibrium Matsubara formalism.

\subsection{Matsubara Green's functions and Gor'kov equations \label{sec:A}}

We consider conduction electrons in the grand canonical ensemble described by Hamiltonian $\hat{\cal H}$ with static electromagnetic fields,
which are expressed here in terms of the static scalar potential $\Phi({\bm r})$ and vector potential ${\bm A}({\bm r})$
as ${\bm E}({\bm r})=-{\bm\nabla}\Phi({\bm r})$ and ${\bm B}({\bm r})={\bm\nabla}\times{\bm A}({\bm r})$.
Let us distinguish the creation and annihilation operators for electrons with integer subscripts $i=1,2$ as
$\hat\psi_1(\xi)\equiv \hat\psi(\xi)$ and 
$\hat\psi_2(\xi)\equiv \hat\psi^\dagger(\xi)$,\cite{KitaText}
where $\xi \equiv ({\bm r}, \alpha)$ with ${\bm r}$ and $\alpha$ denoting the space and spin coordinates, respectively.
Next, we introduce their Heisenberg representations by
$\hat\psi_i(1)\equiv e^{{\tau_1}\hat{\cal H}}\hat\psi_i(\xi_1)e^{-{\tau_1}\hat{\cal H}}$,
where the argument $1$ in the round brackets denotes $1 \equiv (\xi_1,\tau_1)$, 
and the variable $\tau_1$ lies in $0 \le \tau_1 \le 1 / k_{\rm B} T$ 
with $k_{\rm B}$ and $T$ denoting the Boltzmann constant and temperature, respectively. 
Using them, we introduce the Matsubara Green's function:
\begin{equation}
G_{ij} (1,2) \equiv - \langle T_\tau \hat{\psi}_i (1) \hat{\psi}_{3-j} (2) \rangle,
\end{equation}
where $T_\tau$ is the ``time''-ordering operator and $\langle \cdots \rangle$ denotes the grand-canonical average \cite{AGD}.
It can be expanded as
\begin{equation}
G_{ij} (1,2) = k_{\rm B} T \sum_{n=-\infty}^\infty G_{ij} (\xi_1,\xi_2; \varepsilon_n) e^{- i \varepsilon_n (\tau_1-\tau_2)},
\end{equation}
where $\varepsilon_n = (2n + 1) \pi k_{\rm B} T$ is the fermion Matsubara energy $(n=0, \pm 1, \ldots )$.
Separating the spin variable $\alpha = \uparrow, \downarrow$ from $\xi = ({\bm r}, \alpha)$, 
we introduce a new notation for each $G_{ij}$ 
as 
\begin{subequations}
\begin{align}
G_{11} (\xi_1,\xi_2; \varepsilon_n) &= G_{\alpha_1, \alpha_2} ({\bm r}_1, {\bm r}_2; \varepsilon_n), \\
G_{12} (\xi_1,\xi_2; \varepsilon_n) &= F_{\alpha_1, \alpha_2} ({\bm r}_1, {\bm r}_2; \varepsilon_n), \\
G_{21} (\xi_1,\xi_2; \varepsilon_n) &= - \bar{F}_{\alpha_1, \alpha_2} ({\bm r}_1, {\bm r}_2; \varepsilon_n), \\
G_{22} (\xi_1,\xi_2; \varepsilon_n) &= - \bar{G}_{\alpha_1, \alpha_2} ({\bm r}_1, {\bm r}_2; \varepsilon_n). 
\end{align} 
\end{subequations}
Subsequently, we express the spin degrees of freedom as the $2 \times 2$ matrix
\begin{align}
\underline{G} ({\bm r}_1, {\bm r}_2 ; \varepsilon_n) &\equiv
\begin{bmatrix}
G_{\uparrow \uparrow} ({\bm r}_1, {\bm r}_2 ; \varepsilon_n) & G_{\uparrow \downarrow} ({\bm r}_1, {\bm r}_2 ; \varepsilon_n) \\
G_{\downarrow \uparrow} ({\bm r}_1, {\bm r}_2 ; \varepsilon_n) & G_{\downarrow \downarrow} ({\bm r}_1, {\bm r}_2 ; \varepsilon_n)
\end{bmatrix}.
\end{align}
In matrix notation, $\underline{G}$ and $\underline{F}$ satisfy the following symmetry relations: \cite{KitaText}
\begin{subequations}
\begin{align}
\underline{G} ({\bm r}_1, {\bm r}_2 ; \varepsilon_n) &= \underline{G}^\dagger ({\bm r}_1, {\bm r}_2 ; - \varepsilon_n) = \underline{G}^{\rm T} ({\bm r}_1, {\bm r}_2 ; - \varepsilon_n), \\
\underline{F} ({\bm r}_1, {\bm r}_2 ; \varepsilon_n) &= - \underline{F}^\dagger ({\bm r}_1, {\bm r}_2 ; - \varepsilon_n) = - \underline{F}^{\rm T} ({\bm r}_1, {\bm r}_2 ; - \varepsilon_n),
\end{align}
\end{subequations}
where $^\dagger$ and $^{\rm T}$ denote the Hermitian conjugate and transpose, respectively.
It follows from these symmetry relations that $\underline{\bar{G}} ({\bm r}_1, {\bm r}_2 ; \varepsilon_n) = \underline{G}^* ({\bm r}_1, {\bm r}_2 ; \varepsilon_n)$
and $\underline{\bar{F}} ({\bm r}_1, {\bm r}_2 ; \varepsilon_n) = \underline{F}^* ({\bm r}_1, {\bm r}_2 ; \varepsilon_n)$ hold,
where superscript $^*$ denotes the complex conjugate.
Using $\underline{G}$ and $\underline{F}$, we define a $4 \times 4$ Nambu matrix by
\begin{align}
\hat{G} ({\bm r}_1, {\bm r}_2 ; \varepsilon_n) &\equiv
\begin{bmatrix}
\underline{G} ({\bm r}_1, {\bm r}_2 ; \varepsilon_n) & \underline{F} ({\bm r}_1, {\bm r}_2 ; \varepsilon_n) \\
- \underline{F}^* ({\bm r}_1, {\bm r}_2 ; \varepsilon_n) & - \underline{G}^* ({\bm r}_1, {\bm r}_2 ; \varepsilon_n)
\end{bmatrix}.
\label{NambuGreen'sFunctions}
\end{align}
In the mean-field approximation, they satisfy the Gor'kov equations: \cite{Gor'kov,KitaText}
\begin{align}
&
\begin{bmatrix}
( i \varepsilon_n - \hat{\mathcal{K}}_1 ) \underline{\sigma}_0  & \underline{0} \\
\underline{0} & ( i \varepsilon_n + \hat{\mathcal{K}}_1^* ) \underline{\sigma}_0
\end{bmatrix}
\hat{G} ({\bm r}_1, {\bm r}_2 ; \varepsilon_n) \notag \\
&- \int d^3 r_3 \hat{\mathcal{U}}_{\rm BdG} ({\bm r}_1, {\bm r}_3 ) \hat{G} ({\bm r}_3, {\bm r}_2 ; \varepsilon_n) = \hat{\delta} ({\bm r}_1-{\bm r}_2), 
\label{Gor'kovEq}
\end{align}
where $\underline{\sigma}_0$ and  $\underline{0}$ denote the $2 \times 2$ unit and zero matrices, respectively.
Operator $\hat{\mathcal{K}}_1$ is defined by
\begin{equation}
\hat{\mathcal{K}}_1 \equiv \frac{1}{2 m} \left[ - i \hbar \frac{\partial}{\partial {\bm r}_1} - e {\bm A} ({\bm r}_1) \right]^2 + e \Phi ({\bm r}_1) - \mu,
\end{equation}
where $m$ is the electron mass, $e < 0$ is the electron charge, and $\mu$ is the chemical potential. 
Matrix $\hat{\mathcal{U}}_{\rm BdG} ({\bm r}_1, {\bm r}_3 )$ denotes 
\begin{align}
\hat{\mathcal{U}}_{\rm BdG} ({\bm r}_1, {\bm r}_2) &\equiv
\begin{bmatrix}
\underline{\mathcal{U}}_{\rm HF} ({\bm r}_1, {\bm r}_2) & \underline{\Delta} ({\bm r}_1, {\bm r}_2) \\
- \underline{\Delta}^* ({\bm r}_1, {\bm r}_2) & - \underline{\mathcal{U}}_{\rm HF}^* ({\bm r}_1, {\bm r}_2)
\end{bmatrix},
\end{align}
where $\underline{\mathcal{U}}_{\rm HF}$ is the Hartree-Fock potential and $\underline{\Delta}$ is the pair potential \cite{KitaText}.
Finally, matrix $\hat{\delta}$ on the right-hand side of Eq.\ (\ref{Gor'kovEq}) is defined by 
\begin{align}
\hat{\delta} ({\bm r}_1- {\bm r}_2) \equiv
\begin{bmatrix}
\delta ({\bm r}_1- {\bm r}_2) \underline{\sigma}_0 & \underline{0} \\
\underline{0} & \delta ({\bm r}_1-{\bm r}_2) \underline{\sigma}_0
\end{bmatrix}. 
\end{align}

Equation (\ref{Gor'kovEq}) is invariant through the gauge transformation in terms of a continuously differentiable function $\chi ({\bm r})$ \cite{KitaText}, 
\begin{subequations}
\label{GaugeTrans}
\begin{align}
{\bm A} ({\bm r}_1) &= {\bm A}' ({\bm r}_1) + \frac{\partial \chi ({\bm r}_1)}{\partial {\bm r}_1}, \\
\hat{G} ({\bm r}_1, {\bm r}_2 ; \varepsilon_n) &= \hat{\Theta} ({\bm r}_1) \hat{G}' ({\bm r}_1, {\bm r}_2 ; \varepsilon_n) \hat{\Theta}^* ({\bm r}_2), \\
\hat{\mathcal{U}}_{\rm BdG} ({\bm r}_1, {\bm r}_2) &= \hat{\Theta} ({\bm r}_1) \hat{\mathcal{U}}_{\rm BdG}' ({\bm r}_1, {\bm r}_2) \hat{\Theta}^* ({\bm r}_2), 
\end{align}
\end{subequations}
where matrix $\hat{\Theta}$ is defined by 
\begin{align}
\hat{\Theta} ({\bm r}_1) \equiv
\begin{bmatrix}
\underline{\sigma}_0 e^{i e \chi ({\bm r}_1) / \hbar} & \underline{0} \\
\underline{0} & \underline{\sigma}_0 e^{- i e \chi ({\bm r}_1) / \hbar}
\end{bmatrix}. 
\end{align}
%

\subsection{Gauge-covariant Wigner transform \label{sec:B}}

The original Wigner transform \cite{Wigner} breaks the gauge invariance with respect to the center-of-mass coordinate
when applied to the Green's functions of charged systems. 
To remove this drawback, we introduce the gauge-covariant Wigner transform for the Green's functions \cite{KitaText,Kita01}. 
First, we introduce the line integral 
\begin{equation}
I ({\bm r}_1, {\bm r}_2) \equiv \frac{e}{\hbar} \int_{{\bm r}_2}^{{\bm r}_1} {\bm A} ({\bm s}) \cdot d {\bm s}, 
\label{I}
\end{equation}
where ${\bm s}$ denotes a straight-line path from ${\bm r}_2$ to ${\bm r}_1$. 
Next, we define matrix $\hat{\Gamma}$ by 
\begin{align}
\hat{\Gamma} ({\bm r}_1, {\bm r}_2) &\equiv
\begin{bmatrix}
\underline{\sigma}_0 e^{i I ({\bm r}_1, {\bm r}_2)} & \underline{0} \\
\underline{0} & \underline{\sigma}_0 e^{- i I ({\bm r}_1, {\bm r}_2)}
\end{bmatrix}. 
\end{align}
Now, the gauge-covariant Wigner transform for the Green's functions Eq.\ (\ref{NambuGreen'sFunctions}) is defined by
\begin{subequations}
\label{GIWT}
\begin{align}
&\hat{G} (\varepsilon_n, {\bm p}, {\bm r}_{12}) \notag \\
&\equiv \int d^3 \bar{r}_{12} e^{- i {\bm p} \cdot \bar{{\bm r}}_{12} / \hbar} \hat{\Gamma} ({\bm r}_{12}, {\bm r}_1) \hat{G} ({\bm r}_1, {\bm r}_2 ; \varepsilon_n) \hat{\Gamma} ({\bm r}_2, {\bm r}_{12}) \notag \\
&\equiv
\begin{bmatrix}
\underline{G} (\varepsilon_n, {\bm p}, {\bm r}_{12}) & \underline{F} (\varepsilon_n, {\bm p}, {\bm r}_{12}) \\
- \underline{F}^* (\varepsilon_n, - {\bm p}, {\bm r}_{12}) & - \underline{G}^* (\varepsilon_n, - {\bm p}, {\bm r}_{12})
\end{bmatrix}, 
\label{GIWT1}
\end{align}
with ${\bm r}_{12} \equiv ({\bm r}_1 + {\bm r}_2) / 2$ and $\bar{\bm r}_{12} \equiv {\bm r}_1 - {\bm r}_2$, the inverse of which is given by
\begin{align}
&\hat{G} ({\bm r}_1, {\bm r}_2 ; \varepsilon_n) \notag \\
& \ \ \ = \hat{\Gamma} ( {\bm r}_1,{\bm r}_{12}) \int \frac{d^3 p}{(2 \pi \hbar)^3} e^{i {\bm p} \cdot \bar{{\bm r}}_{12} / \hbar} \hat{G} (\varepsilon_n, {\bm p}, {\bm r}_{12}) 
\hat{\Gamma} ({\bm r}_{12},{\bm r}_2). 
\label{GIWT2}
\end{align}
\end{subequations}
It can be shown easily that $\hat{G} (\varepsilon_n, {\bm p}, {\bm r}_{12})$ changes under the gauge transformation in Eq.\ (\ref{GaugeTrans}) to 
\begin{equation}
\hat{G} (\varepsilon_n, {\bm p}, {\bm r}_{12}) = \hat{\Theta} ({\bm r}_{12}) \hat{G}' (\varepsilon_n, {\bm p}, {\bm r}_{12}) \hat{\Theta}^* ({\bm r}_{12}).
\end{equation}
Thus, only the center-of-mass coordinate is relevant to the variation of $\hat{G} (\varepsilon_n, {\bm p}, {\bm r}_{12})$ under the gauge transformation.

\subsection{Derivation of augmented quasiclassical equations \label{sec:C}}

With these preliminaries, we derive the augmented quasiclassical equations in the Matsubara formalism 
following the procedure in Ref. \ \citen{Kita01} for the Keldysh formalism. 

Let us introduce the functions 
\begin{subequations}
\begin{align}
\mathcal{E}_1 (u) &\equiv \int_0^1 d \eta e^{\eta u} = \frac{e^{u} - 1}{u}, \\
\mathcal{E}_2 (u) &\equiv \int_0^1 d \eta \int_0^\eta d \zeta e^{\zeta u} = \frac{e^{u} - 1 - u}{u^2}.
\end{align}
\end{subequations}
The line integral in Eq.\ (\ref{I}) and its partial derivatives are expressible in terms of these functions as 
\begin{equation}
I ({\bm r}_1,{\bm r}_{12}) =  \frac{e}{\hbar} \mathcal{E}_1 \left( \frac{\bar{\bm r}_{12}}{2} \cdot \frac{\partial}{\partial {\bm r}_{12}} \right) \frac{\bar{\bm r}_{12}}{2} \cdot {\bm A} ({\bm r}_{12}), \label{I2} 
\end{equation}
\begin{subequations}
\begin{align}
&\frac{\partial}{\partial {\bm r}_1} I ({\bm r}_1,{\bm r}_{12}) =  \frac{e}{\hbar} {\bm A} ({\bm r}_1) - \frac{e}{2 \hbar} {\bm A} ({\bm r}_{12}) \notag \\
& \ \ \ - \frac{e}{4 \hbar} \left[ 2 \mathcal{E}_1 \left( \frac{\bar{\bm r}_{12}}{2} \cdot \frac{\partial}{\partial {\bm r}_{12}} \right) - \mathcal{E}_2 \left( \frac{\bar{\bm r}_{12}}{2} \cdot \frac{\partial}{\partial {\bm r}_{12}} \right) \right]  \left[ {\bm B} ({\bm r}_{12}) \times \bar{\bm r}_{12} \right], \\
&\frac{\partial}{\partial {\bm r}_1} I ({\bm r}_{12}, {\bm r}_2) =  \frac{e}{2 \hbar} {\bm A} ({\bm r}_{12}) - \frac{e}{4 \hbar} \mathcal{E}_2 \left( \frac{\bar{\bm r}_{12}}{2} \cdot \frac{\partial}{\partial {\bm r}_{12}} \right) \left[ {\bm B} ({\bm r}_{12}) \times \bar{\bm r}_{12} \right].
\end{align}
\end{subequations}
Let us substitute Eq.\ (\ref{GIWT2}) into Eq.\ (\ref{Gor'kovEq}).
Then, the 
kinetic-energy terms can be transformed into 
\begin{subequations}
\label{KGF}
\begin{align}
&\hat{\mathcal{K}}_1 \underline{G} ({\bm r}_1, {\bm r}_2 ; \varepsilon_n) \approx e^{- i I ({\bm r}_{12}, {\bm r}_1)} e^{i I ({\bm r}_{12}, {\bm r}_2)} \int \frac{d^3 p}{(2 \pi \hbar)^3} e^{i {\bm p} \cdot \bar{{\bm r}}_{12} / \hbar} \notag \\
& \ \ \ \times \bigg\{ \xi_p + e \Phi ({\bm r}_{12}) - \frac{i \hbar}{2} {\bm v} \cdot \frac{\partial}{\partial {\bm r}_{12}} - \frac{i \hbar}{2} e {\bm E} ({\bm r}_{12}) \cdot {\bm \partial}_{\bm p} \notag \\
& \ \ \ - \frac{i \hbar}{2} e {\bm v} \cdot \left[ {\bm B} ({\bm r}_{12}) \times {\bm \partial}_{\bm p} \right] \bigg\} \underline{G} (\varepsilon_n, {\bm p}, {\bm r}_{12}), 
\end{align}
\begin{align}
&\hat{\mathcal{K}}_1 \underline{F} ({\bm r}_1, {\bm r}_2 ; \varepsilon_n) \approx e^{- i I ({\bm r}_{12}, {\bm r}_1)} e^{- i I ({\bm r}_{12}, {\bm r}_2)} \int \frac{d^3 p}{(2 \pi \hbar)^3} e^{i {\bm p} \cdot \bar{{\bm r}}_{12} / \hbar} \notag \\
& \ \ \ \times \bigg\{ \xi_p + e \Phi ({\bm r}_{12}) - \frac{i \hbar}{2} {\bm v} \cdot \frac{\partial}{\partial {\bm r}_{12}} - e {\bm v} \cdot {\bm A} ({\bm r}_{12}) \notag \\
& \ \ \ - \frac{i \hbar}{2} e {\bm E} ({\bm r}_{12}) \cdot {\bm \partial}_{\bm p} - \frac{i \hbar}{4} e {\bm v} \cdot \left[ {\bm B} ({\bm r}_{12}) \times {\bm \partial}_{\bm p} \right] \bigg\} \underline{F} (\varepsilon_n, {\bm p}, {\bm r}_{12}),
\end{align}
\begin{align}
&\hat{\mathcal{K}}_1^* \underline{F}^* ({\bm r}_1, {\bm r}_2 ; \varepsilon_n) \approx e^{i I ({\bm r}_{12}, {\bm r}_1)} e^{i I ({\bm r}_{12}, {\bm r}_2)} \int \frac{d^3 p}{(2 \pi \hbar)^3} e^{i {\bm p} \cdot \bar{{\bm r}}_{12} / \hbar} \notag \\
& \ \ \ \times \bigg\{ \xi_p + e \Phi ({\bm r}_{12}) - \frac{i \hbar}{2} {\bm v} \cdot \frac{\partial}{\partial {\bm r}_{12}} + e {\bm v} \cdot {\bm A} ({\bm r}_{12}) \notag \\
& \ \ \  - \frac{i \hbar}{2} e {\bm E} ({\bm r}_{12}) \cdot {\bm \partial}_{\bm p} + \frac{i \hbar}{4} e {\bm v} \cdot \left[ {\bm B} ({\bm r}_{12}) \times {\bm \partial}_{\bm p} \right] \bigg\} \notag \\
& \ \ \ \times \underline{F}^* (\varepsilon_n, - {\bm p}, {\bm r}_{12}), 
\end{align}
\begin{align}
&\hat{\mathcal{K}}_1^* \underline{G}^* ({\bm r}_1, {\bm r}_2 ; \varepsilon_n) \approx e^{i I ({\bm r}_{12}, {\bm r}_1)} e^{- i I ({\bm r}_{12}, {\bm r}_2)} \int \frac{d^3 p}{(2 \pi \hbar)^3} e^{i {\bm p} \cdot \bar{{\bm r}}_{12} / \hbar} \notag \\
& \ \ \ \times \bigg\{ \xi_p + e \Phi ({\bm r}_{12}) - \frac{i \hbar}{2} {\bm v} \cdot \frac{\partial}{\partial {\bm r}_{12}} - \frac{i \hbar}{2} e {\bm E} ({\bm r}_{12}) \cdot {\bm \partial}_{\bm p} \notag \\
& \ \ \ + \frac{i \hbar}{2} e {\bm v} \cdot \left[ {\bm B} ({\bm r}_{12}) \times {\bm \partial}_{\bm p} \right] \bigg\} \underline{G}^* (\varepsilon_n, - {\bm p}, {\bm r}_{12}), 
\end{align}
\end{subequations}
where $\xi_p \equiv p^2 / 2 m - \mu$,
and ${\bm \partial}$ is defined by
\begin{subequations}
\begin{align}
{\bm \partial} \equiv \left\{ \begin{array}{ll} {\bm \nabla} &  {\rm on} \  \underline{G} \ {\rm or} \ \underline{G}^* \\
\displaystyle {\bm \nabla} - i \frac{2 e {\bm A}}{\hbar} &  {\rm on} \ \underline{F} \\
\displaystyle  {\bm \nabla} + i \frac{2 e {\bm A}}{\hbar} & {\rm on} \ \underline{F}^*
\end{array}\right. .
\end{align} 
\end{subequations}
The following approximations have been adopted in deriving Eq.\ (\ref{KGF}): 
(i) We have neglected spatial derivatives of both ${\bm E}$ and ${\bm B}$, which amounts to setting
 $\mathcal{E}_1 \to 1$ and $\mathcal{E}_2 \to 1/2$. 
(ii) We also have neglected terms second-order in ${\bm \partial}_{{\bm r}_{12}}$, ${\bm E}$, and ${\bm B}$. 
(iii) We have expanded $\Phi$ around ${\bm r}_{12}$ up to the first order in $\bar{\bm r}_{12}$ as 
$\Phi ({\bm r}_1) \approx \Phi ({\bm r}_{12}) - {\bm E} ({\bm r}_{12}) \cdot \bar{\bm r}_{12} / 2$. 
By these procedures, we obtain the Gor'kov equations in the Wigner representation,
\begin{align}
&i \varepsilon_n \hat{G} (\varepsilon_n, {\bm p}, {\bm r}) - \left[ \xi_p + e \Phi ({\bm r}) - \frac{i \hbar}{2} {\bm v} \cdot {\bm \partial} \right] \hat{\tau}_3 \hat{G} (\varepsilon_n, {\bm p}, {\bm r}) \notag \\
& \ \ \ + \frac{i \hbar}{2} e {\bm E} ({\bm r}) \cdot {\bm \partial}_{\bm p} \hat{\tau}_3 \hat{G} (\varepsilon_n, {\bm p}, {\bm r}) \notag \\
& \ \ \ + \frac{i \hbar}{8} e {\bm v} \cdot \left[ {\bm B} ({\bm r}) \times {\bm \partial}_{\bm p} \right] \left[ 3 \hat{G} (\varepsilon_n, {\bm p}, {\bm r}) + \hat{\tau}_3 \hat{G} (\varepsilon_n, {\bm p}, {\bm r}) \hat{\tau}_3 \right] \notag \\
& \ \ \ - \hat{\mathcal{U}}_{\rm BdG} ({\bm p}, {\bm r}) \hat{G} (\varepsilon_n, {\bm p}, {\bm r}) = \hat{1}, 
\label{LGEq}
\end{align}
where $\hat{\tau}_3$ is defined by
\begin{align}
\hat{\tau}_3 \equiv
\begin{bmatrix}
\underline{\sigma}_0 & \underline{0} \\
\underline{0} & - \underline{\sigma}_0
\end{bmatrix}, 
\end{align}
and $\hat{1}$ denotes the $4 \times 4$ unit matrix. 
We take the Hermitian conjugate of Eq.\ (\ref{LGEq}), use the symmetries 
$\hat{\mathcal{U}}_{\rm BdG}^\dagger ({\bm p}, {\bm r}) = \hat{\mathcal{U}}_{\rm BdG} ({\bm p}, {\bm r})$ 
and $\hat{G}^\dagger (\varepsilon_n, {\bm p}, {\bm r}) = \hat{G} ( - \varepsilon_n, {\bm p}, {\bm r})$, 
and set $\varepsilon_n \to - \varepsilon_n$ to obtain 
\begin{align}
&i \varepsilon_n \hat{G} (\varepsilon_n, {\bm p}, {\bm r}) - \left[ \xi_p + e \Phi ({\bm r}) + \frac{i \hbar}{2} {\bm v} \cdot {\bm \partial} \right] \hat{G} (\varepsilon_n, {\bm p}, {\bm r}) \hat{\tau}_3 \notag \\
& \ \ \ - \frac{i \hbar}{2} e {\bm E} ({\bm r}) \cdot {\bm \partial}_{\bm p} \hat{G} (\varepsilon_n, {\bm p}, {\bm r}) \hat{\tau}_3 \notag \\
& \ \ \ - \frac{i \hbar}{8} e {\bm v} \cdot \left[ {\bm B} ({\bm r}) \times {\bm \partial}_{\bm p} \right] \left[ 3 \hat{G} (\varepsilon_n, {\bm p}, {\bm r}) + \hat{\tau}_3 \hat{G} (\varepsilon_n, {\bm p}, {\bm r}) \hat{\tau}_3 \right] \notag \\
& \ \ \ - \hat{G} (\varepsilon_n, {\bm p}, {\bm r}) \hat{\mathcal{U}}_{\rm BdG} ({\bm p}, {\bm r}) = \hat{1} .
\label{RGEq}
\end{align}
Equations (\ref{LGEq}) and (\ref{RGEq}) are referred to as the left and right Gor'kov equations, respectively. 
Now, we operate $\hat{\tau}_3$ from the left and right sides of Eq.\ (\ref{RGEq})
and subtract the resulting equation from Eq.\ (\ref{LGEq}). 
We thereby obtain  
\begin{align}
&\left[ i \varepsilon_n \hat{\tau}_3 - \hat{\mathcal{U}}_{\rm BdG} ({\bm p}, {\bm r}) \hat{\tau}_3, \hat{\tau}_3 \hat{G} (\varepsilon_n, {\bm p}, {\bm r}) \right] \notag \\
& \ \ \ + i \hbar {\bm v} \cdot {\bm \partial} \hat{\tau}_3 \hat{G} (\varepsilon_n, {\bm p}, {\bm r}) + i \hbar e {\bm E} ({\bm r}) \cdot {\bm \partial}_{\bm p} \hat{\tau}_3 \hat{G} (\varepsilon_n, {\bm p}, {\bm r}) \notag \\
& \ \ \ + \frac{i \hbar}{2} e {\bm v} \cdot \left[ {\bm B} ({\bm r}) \times {\bm \partial}_{\bm p} \right] \left\{ \hat{\tau}_3, \hat{\tau}_3 \hat{G} (\varepsilon_n, {\bm p}, {\bm r}) \right\} = \hat{0}, 
\end{align}
with $[ \hat{a}, \hat{b} ] \equiv \hat{a} \hat{b} - \hat{b} \hat{a}$ and $\{ \hat{a}, \hat{b} \} \equiv \hat{a} \hat{b} + \hat{b} \hat{a}$.

Finally, we perform integration over $\xi_p$ neglecting all the $\xi_p$ dependences except those in $\hat G$.
To this end, we introduce  the quasiclassical Green's functions: 
\begin{align}
\hat{g}  (\varepsilon_n, {\bm p}_{\rm F}, {\bm r}) &\equiv {\rm P} \int_{- \infty}^\infty \frac{d \xi_p}{\pi} \hat{\tau}_3 i \hat{G} (\varepsilon_n, {\bm p}, {\bm r}) \notag \\
&\equiv
\begin{bmatrix}
\underline{g} (\varepsilon_n, {\bm p}_{\rm F}, {\bm r}) & - i \underline{f} (\varepsilon_n, {\bm p}_{\rm F}, {\bm r}) \\
- i \underline{f}^* (\varepsilon_n, - {\bm p}_{\rm F}, {\bm r}) & - \underline{g}^* (\varepsilon_n, - {\bm p}_{\rm F}, {\bm r})
\end{bmatrix}, 
\end{align}
where P denotes the principal value. 
We also carry out the following procedures to obtain the final equations:
(i) Rewrite ${\mbox{\boldmath$\partial$}}_{{\bf p}}\!=\!
{\mbox{\boldmath$\partial$}}_{{\bf p}_{\parallel}}\!+\!
{\bm v}_{\rm F}(\partial/\partial \xi_p)$
with ${\bf p}_{\parallel}$ the component on the energy surface of $\xi_p={\rm constant}$.
(ii) Make use of
${\bm v}_{\rm F}\times{\mbox{\boldmath$\partial$}}_{{\bf p}_{\parallel}}
= {\bm v}_{\rm F}\times{\mbox{\boldmath$\partial$}}_{{\bf p}}$ 
and
\begin{align}
{\rm P} \int_{-\infty}^{\infty}
\! d \xi_p \,  \frac{\partial^{m}}{\partial \xi_p^{m}} \hat{\tau}_3 i \hat{G} (\varepsilon_n, {\bm p}, {\bm r})=0 ,\hspace{5mm}(m=1,2,\cdots).
\notag
\end{align}
(iii) Neglect the term
${\bm E}\cdot {\mbox{\boldmath$\partial$}}_{{\bf p}_{\parallel}}$ 
because it is second-order in the quasiclassical parameter $\delta \equiv \hbar / \langle p_{\rm F} \rangle_{\rm F} \xi_0 \ll 1$ \cite{Kita01,Kita09},
where $\xi_0$ is the coherence length defined in terms of
the zero-temperature energy gap $\langle \Delta_0\rangle_{\rm F}$ at $B=0$ by $\xi_0\equiv \hbar \langle v_{\rm F}\rangle_{\rm F}/\langle \Delta_0\rangle_{\rm F}$.
We thereby obtain the augmented quasiclassical equations in the Matsubara formalism as 
\begin{align}
&\left[ i \varepsilon_n \hat{\tau}_3 - \hat{\mathcal{U}}_{\rm BdG} ({\bm p}_{\rm F}, {\bm r}) \hat{\tau}_3, \hat{g} (\varepsilon_n, {\bm p}_{\rm F}, {\bm r}) \right] 
+ i \hbar {\bm v}_{\rm F} \cdot {\bm \partial} \hat{g} (\varepsilon_n, {\bm p}_{\rm F}, {\bm r}) 
\notag \\
& \ \ \ + \frac{i \hbar}{2} e {\bm v}_{\rm F} \cdot \left[ {\bm B} ({\bm r}) \times {\bm \partial}_{{\bm p}_{\rm F}} \right] \left\{ \hat{\tau}_3, \hat{g} (\varepsilon_n, {\bm p}_{\rm F}, {\bm r}) \right\} = \hat{0}.
\end{align}
Thus, the electric field is absent from the equations in the Matsubara formalism unlike those in the Keldysh formalism.

Now, we consider the weak-coupling case and include the effects of impurity scatterings in the self-consistent Born approximation by\cite{KitaText}
 $\hat{\mathcal{U}}_{\rm BdG} ({\bm p}_{\rm F}, {\bm r}) \!\to\! \hat{\Delta} ({\bm p}_{\rm F}, {\bm r}) \!+\! \hat{\sigma}_{\rm imp} (\varepsilon_n, {\bm r})$. 
The pair potentials $\hat{\Delta} ({\bm p}_{\rm F}, {\bm r})$ and impurity self-energy $\hat{\sigma}_{\rm imp} (\varepsilon_n, {\bm r})$ are given explicitly by
\begin{subequations}
\begin{align}
\hat{\Delta} ({\bm p}_{\rm F}, {\bm r}) &\equiv
\begin{bmatrix}
\underline{0} & \underline{\Delta} ({\bm p}_{\rm F}, {\bm r}) \\
- \underline{\Delta}^* (- {\bm p}_{\rm F}, {\bm r}) & \underline{0}
\end{bmatrix}, \\
\hat{\sigma}_{\rm imp} (\varepsilon_n, {\bm r}) &\equiv - i \frac{\hbar}{2 \tau} \langle \hat{g} (\varepsilon_n, {\bm p}_{\rm F}, {\bm r}) \rangle_{\rm F} \hat{\tau}_3,
\end{align}
\end{subequations}
where $\tau$ is the relaxation time and $\langle \cdots \rangle_{\rm F}$ denotes the Fermi surface average with $\langle 1 \rangle_{\rm F} = 1$.
The augmented quasiclassical equations in the Matsubara formalism are then given by 
\begin{align}
&\left[ i \varepsilon_n \hat{\tau}_3 - \hat{\Delta} \hat{\tau}_3 - \hat{\sigma}_{\rm imp} \hat{\tau}_3, \hat{g} \right] \notag \\
& \ \ \ + i \hbar {\bm v}_{\rm F} \cdot {\bm \partial} \hat{g} + \frac{i \hbar}{2} e ({\bm v}_{\rm F} \times {\bm B}) \cdot \frac{\partial}{\partial {\bm p}_{\rm F}}  \left\{ \hat{\tau}_3, \hat{g} \right\} = \hat{0}. 
\label{QCE-Matsubara}
\end{align}
Matrices $\hat{g}$ and $\hat{\Delta}$ can be written as \cite{KitaText}
\begin{align}
\hat{g}  \equiv
\begin{bmatrix}
\vspace{1mm}
\underline{g} & - i \underline{f} \\
- i \underline{\bar{f}} & - \underline{\bar{g}}
\end{bmatrix}, \hspace{10mm}
\hat{\Delta} \equiv
\begin{bmatrix}
\vspace{1mm}
\underline{0} & \underline{\Delta} \\
- \underline{\bar{\Delta}} & \underline{0}
\end{bmatrix},
\end{align}
where the barred functions are defined generally by $\underline{\bar{g}} (\varepsilon_n, {\bm p}_{\rm F}, {\bm r}) \equiv \underline{g}^* (\varepsilon_n, - {\bm p}_{\rm F}, {\bm r})$.
It is worth pointing out that the same equations result 
in the gauge  ${\bm E}({\bm r}) = - \partial {\bm A}'({\bm r}, t) / \partial t$ and ${\bm B}({\bm r}) = {\bm \nabla} \times {\bm A}'({\bm r}, t)$ with $\Phi'=0$. 
The gauge transformation $(\Phi, {\bm A}) \to (0, {\bm A}')$ is given by 
\begin{subequations}
\begin{align}
\Phi ({\bm r}) &= - \frac{\partial \chi({\bm r}, t)}{\partial t}, \\
{\bm A} ({\bm r}) &= {\bm A}' ({\bm r}, t) + {\bm \nabla} \chi ({\bm r}, t), \\
\underline{g} (\varepsilon_n, {\bm p}_{\rm F}, {\bm r}) &= \underline{g}' (\varepsilon_n, {\bm p}_{\rm F}, {\bm r}), \\
\underline{f} (\varepsilon_n, {\bm p}_{\rm F}, {\bm r}) &= \underline{f}' (\varepsilon_n, {\bm p}_{\rm F}, {\bm r}) e^{2 i e \chi ({\bm r}, t) / \hbar},
\end{align} \label{gauge-transform}%
\end{subequations}
where the continuously differentiable function $\chi ({\bm r}, t)$ is fixed by 
\begin{align}
{\bm \nabla} \chi ({\bm r}, t) = {\bm E} ({\bm r}) t, \ \ \ \frac{\partial {\bm \nabla} \chi ({\bm r}, t)}{\partial t} = - \frac{\partial {\bm A}' ({\bm r}, t)}{\partial t}. \label{chi}
\end{align}
%

\section{Analytic Continuation in Terms of Frequency \label{sec:III}}

Next, we consider the augmented quasiclassical equations in the Keldysh formalism 
and study their connection with Eq.\ (\ref{QCE-Matsubara}). 
It is convenient when describing equilibrium states in the Keldysh formalism to set $\Phi' \!\rightarrow\! 0$ 
and express static electromagnetic fields in terms of only the vector potential ${\bm A}'$ with linear time dependence as
${\bm E}({\bm r}) = - \partial {\bm A}'({\bm r}, t) / \partial t$ and ${\bm B}({\bm r}) = {\bm \nabla} \times {\bm A}'({\bm r}, t)$.
The rationale for this is that the scalar potential $\Phi'$ in the Keldysh formalism always appears in the covariant form $i\hbar\partial/\partial t -2e\Phi'$,\cite{Kita01}
which in the present gauge can be set equal to zero naturally for static situations. 
Thus, we derive the augmented quasiclassical equations in the Keldysh formalism in the static case using the following line integral: 
\begin{equation}
I (\vec{r}_1, \vec{r}_2) \equiv - \frac{e}{\hbar} \int_{\vec{r}_2}^{\vec{r}_1} \vec{A} (\vec{s}) \cdot d \vec{s}, 
\end{equation}
where $\vec{r}_1 \equiv (t_1, {\bm r}_1)$ is the four-vector, $d \vec{s}$ is taken along the straight line, and $\vec{A} ({\bm r}, t)$ is given by
\begin{equation}
\vec{A} ({\bm r}, t) \equiv \left( - \frac{\partial \chi ({\bm r}, t)}{\partial t}, - {\bm A}' ({\bm r}, t) - {\bm \nabla} \chi ({\bm r}, t) \right), 
\end{equation}
where $\chi ({\bm r}, t)$ is also fixed as Eq. \ (\ref{chi}). 
The gauge-covariant Wigner transform for the retarded Green's functions is now given by 
\begin{align}
&\hat{G}^{\rm R} (\varepsilon, {\bm p}, {\bm r}_{12}) \notag \\
&\equiv \int d^3 \bar{r}_{12} d \bar{t}_{12} e^{- i ({\bm p} \cdot \bar{{\bm r}}_{12} - \varepsilon \bar{t}_{12}) / \hbar} \hat{\Gamma} (\vec{r}_{12}, \vec{r}_1) \hat{G}^{\rm R} ({\bm r}_1, {\bm r}_2; \bar{t}_{12}) \hat{\Gamma} (\vec{r}_2, \vec{r}_{12}) \notag \\
&\equiv
\begin{bmatrix}
\underline{G}^{\rm R} (\varepsilon, {\bm p}, {\bm r}_{12}) & \underline{F}^{\rm R} (\varepsilon, {\bm p}, {\bm r}_{12}) \\
- \underline{F}^{{\rm R}*} (- \varepsilon, - {\bm p}, {\bm r}_{12}) & - \underline{G}^{{\rm R}*} (- \varepsilon, - {\bm p}, {\bm r}_{12})
\end{bmatrix}, 
\end{align}
where $t_{12} \equiv (t_1 + t_2)/2$, $\bar{t}_{12} \equiv t_1 - t_2$, and matrix $\hat{\Gamma}$ is defined by 
\begin{align}
\hat{\Gamma} (\vec{r}_1, \vec{r}_2) &\equiv
\begin{bmatrix}
\underline{\sigma}_0 e^{i I (\vec{r}_1, \vec{r}_2)} & \underline{0} \\
\underline{0} & \underline{\sigma}_0 e^{- i I (\vec{r}_1, \vec{r}_2)}
\end{bmatrix}. 
\end{align}

The corresponding
augmented quasiclassical equations for the retarded submatrix $\hat{g}^{\rm R}=\hat{g}^{\rm R}(\varepsilon,{\bm p}_{\rm F},{\bm r})$ 
 are given by \cite{Kita01,Kita09}
\begin{subequations}
\begin{align}
&\left[ \varepsilon \hat{\tau}_3 - \hat{\Delta} \hat{\tau}_3 - \hat{\sigma}_{\rm imp}^{\rm R} \hat{\tau}_3, \hat{g}^{\rm R} \right] + i \hbar {\bm v}_{\rm F} \cdot {\bm \partial} \hat{g}^{\rm R} \notag \\
& \ \ \ + \frac{i \hbar}{2} \left[ e {\bm v}_{\rm F} \cdot {\bm E} \frac{\partial}{\partial \varepsilon} + e ({\bm v}_{\rm F} \times {\bm B}) \cdot \frac{\partial}{\partial {\bm p}_{\rm F}} \right] \left\{ \hat{\tau}_3, \hat{g}^{\rm R} \right\} = \hat{0}, \label{gREq} \\
&\hat{\sigma}_{\rm imp}^{\rm R} \equiv - \frac{i \hbar}{2 \tau} \langle \hat{g}^{\rm R} \rangle_{\rm F} \hat{\tau}_3.
\end{align} \label{QCE-Keldysh}%
\end{subequations}
The quasiclassical Green's function $\hat g^{\rm R}$ is expressible as \cite{Kita01,Kita09} 
\begin{align}
\hat g^{\rm R} =\begin{bmatrix}\vspace{1mm} \underline{g}^{\rm R} & -i \underline{f}^{\rm R}
\\
-i \underline{\bar f}^{\rm R} &  -\underline{\bar g}^{\rm R}
\end{bmatrix},
\end{align}
where each barred $2\times 2$ submatrix is connected generally to its unbarred equivalent as 
$\underline{\bar g}^{\rm R}(\varepsilon,{\bm p}_{\rm F},{\bm r})=\underline{g}^{{\rm R}*}(-\varepsilon,-{\bm p}_{\rm F},{\bm r})$. 
Thus, Eq.\ (\ref{QCE-Keldysh}) manifestly contains an electric-field term, which is absent in Eq.\ (\ref{QCE-Matsubara}), however. 
The issue here is how to perform the analytic continuation between $\hat{g}'$ and $\hat{g}^{\rm R}$ obeying Eqs.\ 
 (\ref{QCE-Matsubara}) and (\ref{QCE-Keldysh}) with different forms. 
 Alternatively, one may depend solely on Eq. \ (\ref{QCE-Keldysh}) and put $\varepsilon \!\to\! i \varepsilon_n$ directly; 
however, this procedure also has a difficulty in how to perform differentiation with respect to $\varepsilon_n$, which has discrete values. 

To find the procedure, we extract the (1,1) and (1,2) submatrix elements from Eq.\ (\ref{QCE-Keldysh}). 
They can be written explicitly as 
\begin{subequations}
\begin{align}
&\hbar {\bm v}_{\rm F} \cdot {\bm \nabla} \underline{g}^{\rm R} + \hbar e {\bm v}_{\rm F} \cdot {\bm E} \frac{\partial \underline{g}^{\rm R}}{\partial \varepsilon} + \hbar e ({\bm v}_{\rm F} \times {\bm B}) \cdot \frac{\partial \underline{g}^{\rm R}}{\partial {\bf p}_{\rm F}}  \notag \\
& \ \ \ - \underline{\Delta} \underline{\bar{f}}^{\rm R} + \underline{f}^{\rm R} \underline{\bar{\Delta}} + \frac{\hbar}{2 \tau} \left( \underline{f}^{\rm R} \langle \underline{\bar{f}}^{\rm R} \rangle_{\rm F} - \langle \underline{f}^{\rm R} \rangle_{\rm F} \underline{\bar{f}}^{\rm R} \right) = \underline{0}, 
\label{g-eq}
\\
&- 2 i \varepsilon \underline{f}^{\rm R} + \hbar {\bm v}_{\rm F} \cdot \left( {\bm \nabla} - i \frac{2 e {\bm A}'}{\hbar} \right) \underline{f}^{\rm R} - \underline{\Delta} \underline{\bar{g}}^{\rm R} - \underline{g}^{\rm R} \underline{\Delta} \notag \\
& \ \ \ + \frac{\hbar}{2 \tau} \left( \langle \underline{g}^{\rm R} \rangle_{\rm F} \underline{f}^{\rm R} - \langle \underline{f}^{\rm R} \rangle_{\rm F} \underline{\bar{g}}^{\rm R} - \underline{g}^{\rm R} \langle \underline{f}^{\rm R} \rangle_{\rm F} + \underline{f}^{\rm R} \langle \underline{\bar{g}}^{\rm R} \rangle_{\rm F} \right)
= \underline{0}.
\end{align}
\end{subequations}
We then write the gradient term in Eq.\ (\ref{g-eq}) together with the electric-field term as
\begin{align}
{\bm\nabla}\underline{g}^{\rm R}+e{\bm E}\frac{\partial \underline{g}^{\rm R}}{\partial\varepsilon} \equiv {\bm\nabla}\underline{\tilde g}^{\rm R} ,
\label{nabla-g^R}
\end{align}
and eliminate $\underline{g}^{\rm R}$ in the two equations in favor of $\underline{\tilde g}^{\rm R}$.
We then use (i) the smallness of the Lorentz term by $\delta \ll 1$ \cite{Kita09}. 
(ii) $\underline{g}^{\rm R}\propto \underline{\sigma}_0$ for the leading order and
(iii) $\bar{g}^{\rm R} - g^{\rm R} = O (\delta)$,
to neglect terms of $O(\delta^2)$.
The procedure yields
\begin{subequations}
\begin{align}
&\hbar {\bm v}_{\rm F} \cdot {\bm \nabla} \underline{\tilde{g}}^{\rm R} + \hbar e ({\bm v}_{\rm F} \times {\bm B}) \cdot \frac{\partial \underline{\tilde{g}}^{\rm R}}{\partial {\bf p}_{\rm F}} \notag \\
& \ \ \ - \underline{\Delta} \underline{\bar{f}}^{\rm R} + \underline{f}^{\rm R} \underline{\bar{\Delta}} + \frac{\hbar}{2 \tau} \left( \underline{f}^{\rm R} \langle \underline{\bar{f}}^{\rm R} \rangle_{\rm F} - \langle \underline{f}^{\rm R} \rangle_{\rm F} \underline{\bar{f}}^{\rm R} \right) = \underline{0}, \\
&- 2 i \varepsilon \underline{f}^{\rm R} + \hbar {\bm v}_{\rm F} \cdot \left( {\bm \nabla} - i \frac{2 e {\bm A}'}{\hbar} \right) \underline{f}^{\rm R} - \underline{\Delta} \underline{\bar{\tilde{g}}}^{\rm R} - \underline{\tilde{g}}^{\rm R} \underline{\Delta} \notag \\
& \ \ \ + \frac{\hbar}{2 \tau} \left( \langle \underline{\tilde{g}}^{\rm R} \rangle_{\rm F} \underline{f}^{\rm R} - \langle \underline{f}^{\rm R} \rangle_{\rm F} \underline{\bar{\tilde{g}}}^{\rm R} - \underline{\tilde{g}}^{\rm R} \langle \underline{f}^{\rm R} \rangle_{\rm F} + \underline{f}^{\rm R} \langle \underline{\bar{\tilde{g}}}^{\rm R} \rangle_{\rm F} \right)
 = \underline{0}.
\end{align}
\label{gtildeREq}
\end{subequations}
These equations are identical in form with those for $(\underline{g}',\underline{f}')$ from 
Eq.\ (\ref{QCE-Matsubara}) transformed by Eq. \ (\ref{gauge-transform}), as can be seen easily.
This implies that we may perform the analytic continuation in terms of $\varepsilon_n>0$ using 
\begin{equation}
\left\{\begin{array}{l}
\vspace{1mm}
\underline{g}' (\varepsilon_n , {\bm p}_{\rm F}, {\bm r}) =  \underline{\tilde{g}}^{\rm R} (i \varepsilon_n, {\bm p}_{\rm F}, {\bm r}) \\
\underline{f}' (\varepsilon_n, {\bm p}_{\rm F}, {\bm r}) =  \underline{f}^{\rm R} (i \varepsilon_n, {\bm p}_{\rm F}, {\bm r})
\end{array}\right. .
\label{AC}
\end{equation}

Accordingly, the expression for the charge density  in the Matsubara formalism needs to be modified. 
To see this, we start from the expression in the Keldysh formalism \cite{SR,ESR}:
\begin{align}
\rho = - \frac{e N (0)}{4} \int_{- \infty}^\infty  {\rm Tr} \langle\, \underline{g}^{\rm K} \rangle_{\rm F}\, d \varepsilon.
\notag
\end{align}
Here, $N (0)$ is the normal density of states per spin and unit volume at the Fermi energy,
Tr denotes the trace in spin space, and
$\underline{g}^{\rm K} = (\underline{g}^{\rm R} - \underline{g}^{\rm A}) \tanh (\varepsilon / 2 k_{\rm B} T)$ in equilibrium
with $\underline{g}^{{\rm A}} \equiv - \underline{\sigma}_3 \underline{g}^{{\rm R} \dagger} \underline{\sigma}_3$,
where $\underline{\sigma}_3$ denotes the third Pauli matrix.
Let us apply the operater ${\bm\nabla}$ to this equation, substitute Eq.\ (\ref{nabla-g^R}), 
and use ${\rm Tr} \,\underline{g}^{\rm K}\rightarrow \pm 4$ for $\varepsilon\rightarrow\pm\infty$
to perform integration with respect to $\varepsilon$ for the electric-field term.
This leads to
\begin{align}
{\bm\nabla}\rho =&\, - \frac{e N (0)}{4}{\bm\nabla}  \int_{- \infty}^\infty  {\rm Tr} 
\langle \, \underline{\tilde{g}}^{\rm R}- \underline{\tilde{g}}^{\rm A} \rangle_{\rm F} \tanh \frac{\varepsilon}{2 k_{\rm B} T}\, d \varepsilon
\notag \\
&\,  +2e^2 N (0){\bm E}.
\notag
\end{align}
Deforming the contour of the above integral towards the imaginary axis 
using the residue theorem, and noting Eq.\ (\ref{AC}), 
we can express the charge density in terms of $\underline{g}(\varepsilon_n, {\bm p}_{\rm F}, {\bm r})$ as
\begin{align}
\rho &= - i  \pi k_{\rm B} T e N (0)  \sum_{n = - \infty}^\infty {\rm Tr} \langle \underline{g} \rangle_{\rm F} - 2 e^2 N (0) \Phi. 
\label{rho} 
\end{align}
This expression is the same as that in Refs.\ \citen{Kopnin,ESR}, and \citen{Eliashberg}.
On the other hand, the formula for the current density has no extra term with ${\bm E}$ because $\langle {\bm v}_{\rm F}\rangle_{\rm F}={\bm 0}$,
and so is the equation for the energy gap \cite{SR,ESR,KitaText}.
This argument is valid even when the impurity self-energy is incorporated. 
This completes our formulation of the augmented quasiclassical equations in the Matsubara formalism.

\section{Equation of Electric Field and Expression for Vortex-Core Charge \label{sec:IV}}

We now solve Eqs.\ (\ref{QCE-Matsubara}) and  (\ref{rho}) for the spin-singlet pairing without spin paramagnetism,
where $\underline{g}$ is expressible as $\underline{g}=g\underline{\sigma}_0$.
As in Ref.\ \citen{Kita09}, we expand $g$ and  $\rho$ formally in $\delta$ as $g=g_0+g_1+\cdots$ and
$\rho=\rho_0+\rho_1+\cdots$, where
$g_0$ and $\rho_0 = 0$ are the solutions of the standard Eilenberger equations \cite{SR,Kopnin}.
We then find that $g_1$ is expressible in terms of $g_0$ as \cite{Kita09}
\begin{equation}
{\bm \nabla} g_1 = - e {\bm B} \times \frac{\partial g_0}{\partial {\bm p}_{\rm F}}. 
\label{nablag1}
\end{equation}
Next, we apply operater ${\bm \nabla}$ to Eq.\ (\ref{rho}), 
substitute Eq.\ (\ref{nablag1}) and Gauss' law $\rho=\epsilon_0 {\bm \nabla} \cdot {\bm E}$ ($\epsilon_0$:  vaccum permittivity), 
and use ${\bm\nabla}\times{\bm E}={\bm 0}$.
We thereby obtain
\begin{equation}
- \lambda_{\rm TF}^2 {\bm \nabla}^2 {\bm E} + {\bm E} = - i  \pi k_{\rm B} T  {\bm B} \times \sum_{n = - \infty}^\infty \left\langle \frac{\partial g_0}{\partial {\bm p}_{\rm F}} \right\rangle_{\rm F}, 
\label{EEq}
\end{equation}
where $\lambda_{\rm TF} \!\equiv\! \sqrt{\epsilon_0/2 e^2 N (0)}$ is the Thomas-Fermi screening length. 
This equation enables us to calculate the electric field and charge density microscopically  
even in the presence of impurity scattering
based on the solution of the standard Eilenberger equations in the Matsubara formalism.

For extreme type-II materials in the clean limit, we can also estimate the vortex-core charge analytically based on Eq.\ (\ref{EEq}) and the charge neutrality condition. 
It follows from Eq.\ (\ref{EEq}) that the electric field outside the core obeys 
\begin{align}
{\bm E} = {\bm B} \times \underline{R}_{\rm H} {\bm j}
\label{E}
\end{align}
with the tensor Hall coefficient \cite{Kita09}
\begin{align}
\underline{R}_{\rm H} \equiv \frac{1}{2 e N(0)} \!\left\langle \frac{\partial}{\partial {\bm p}_{\rm F}} (1\! -\! Y) {\bm v}_{\rm F} \right\rangle_{\!\!{\rm F}} \langle {\bm v}_{\rm F} (1\!-\! Y) {\bm v}_{\rm F}\rangle_{\rm F}^{-1},
\label{R_H}
\end{align} 
where $Y\!=\!Y({\bm p}_{\rm F},T)$ is the Yosida function \cite{Kita09,KitaText}.
Assuming cylindrical symmetry outside the core, 
we can express the flux density and supercurrent as\cite{KitaText}
\begin{align}
B(r)=\frac{\Phi_0}{2\pi\lambda_{\rm L}^2} K_0\left(\frac{r}{\lambda_{\rm L}}\right),\hspace{5mm}
j(r)=\frac{\Phi_0}{2\pi\lambda_{\rm L}^3\mu_0} K_1\left(\frac{r}{\lambda_{\rm L}}\right),
\end{align}
where $K_{0,1}(x)$ are the modified Bessel functions, 
 $\lambda_{\rm L}$ is the London penetration depth
 at finite temperatures,
and $\Phi_0\equiv h/2|e|$ and $\mu_0$ denote the magnetic flux quantum and vacuum permeability, respectively.
Using them in Eq.\ (\ref{E}), we obtain the electric field along the radial direction as 
\begin{align}
E(r)=-\frac{R_{\rm H}\Phi_0^2}{4\pi^2\lambda_{\rm L}^5 \mu_0}K_0\left(\frac{r}{\lambda_{\rm L}}\right) K_1\left(\frac{r}{\lambda_{\rm L}}\right),
\end{align}
where $R_{\rm H}$ denotes  
the diagonal element of $\underline{R}_{\rm H}$.
We then integrate the resulting charge density $\rho=\epsilon_0 {\bm\nabla} \cdot {\bm E}$ 
over $r_{\rm c}\leq r\leq \infty$  with $r_{\rm c}\sim\xi_0$ to estimate the charge accumulated in the outer region per unit length along the flux line, 
which should be equal in magnitude and opposite in sign to that
in $r\leq r_{\rm c}$ due to the charge neutrality condition.
We thereby obtain the following expression for the vortex-core charge within $r \le r_{\rm c}$ per unit length along the flux line:  
\begin{equation}
Q_\lambda = -\frac{\epsilon_0R_{\rm H}\Phi_0^2 r_{\rm c}}{2\pi\lambda_{\rm L}^5 \mu_0} K_0\left(\frac{r_{\rm c}}{\lambda_{\rm L}}\right)K_1\left(\frac{r_{\rm c}}{\lambda_{\rm L}}\right)\!\approx
\frac{e^2R_{\rm H}}{32 \pi \alpha^2\lambda_{\rm L}^4} \ln \frac{r_{\rm c}}{\lambda_{\rm L}}, \label{Q}
\end{equation}
where $\alpha \equiv e^2 / 4 \pi \epsilon_0 \hbar c$ is the fine-structure constant with  $c$ the light velocity, and 
we have used $K_0(x)\approx -\ln x$ and $K_1(x)\approx 1/x$ for $x\equiv r_{\rm c}/\lambda_{\rm L}\ll1$. 
Equation (\ref{Q}) implies that the magnitude of the core charge depends crucially on $\lambda_{\rm L}$.
It also follows from Eq.\ (\ref{R_H}) that both the sign and magnitude of $Q_\lambda$ are strongly affected by 
the curvature of the Fermi surface and may also exhibit substantial temperature dependence in the presence of gap anisotropy
due to the factor $Y\!=\!Y({\bm p}_{\rm F},T)$.

\begin{figure}[t]
        \begin{center}
                \includegraphics[width=0.95\linewidth]{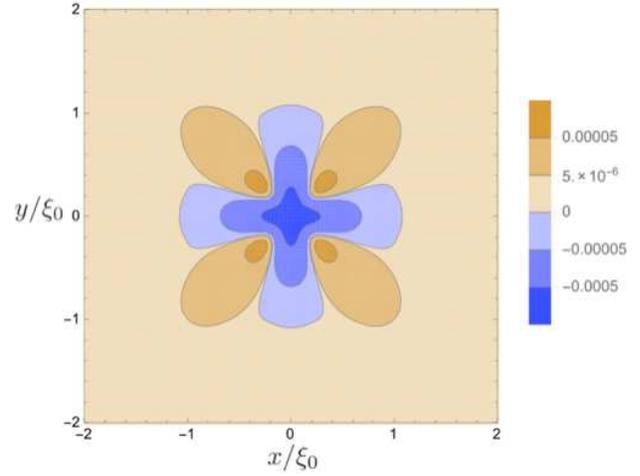}
                \end{center}
\caption{
(Color online) Charge density $\rho ({\bm r})$ at $T = 0.2 T_{\rm c}$ in units of $\rho_0 \equiv \epsilon_0 \Delta_0 / | e | \xi_0^2$ over $-2\xi_0 \!\le\! x, y \!\le\! 2\xi_0$ for
$n=1.95$ with an isotropic holelike Fermi surface. 
}
\label{fig1}
\end{figure}
\begin{figure}[t]
        \begin{center}
                \includegraphics[width=0.95\linewidth]{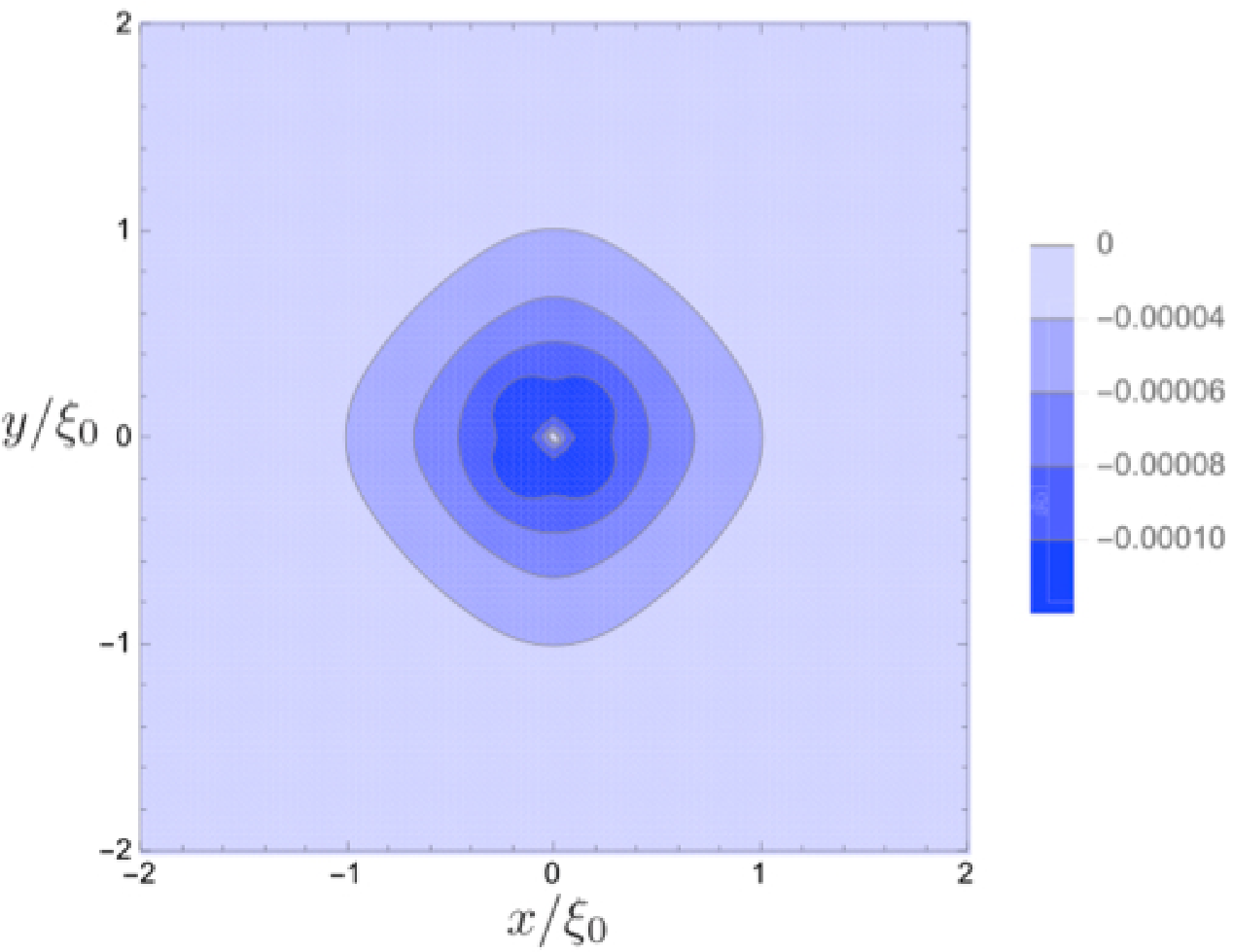}
                \end{center}
\caption{
(Color online) Electric field along the radial direction $E_r ({\bm r})$ at $T = 0.2 T_{\rm c}$ in units of $E_0 \equiv \Delta_0 / | e | \xi_0$ over $-2\xi_0 \!\le\! x, y \!\le\! 2\xi_0$ for
$n=1.95$ with an isotropic holelike Fermi surface. }
\label{fig2}
\end{figure}
\begin{figure}[t]
        \begin{center}
                \includegraphics[width=0.95\linewidth]{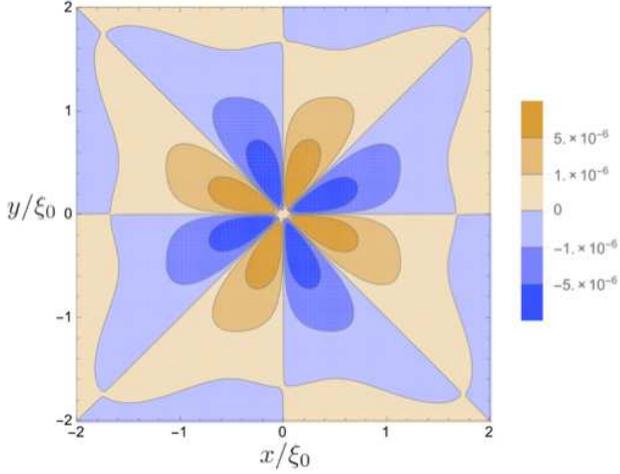}
                \end{center}
\caption{
(Color online) Electric field along the angular direction $E_\varphi ({\bm r})$ at $T = 0.2 T_{\rm c}$ in units of $E_0 \equiv \Delta_0 / | e | \xi_0$ over $-2\xi_0 \!\le\! x, y \!\le\! 2\xi_0$ for
$n=1.95$ with an isotropic holelike Fermi surface. }
\label{fig3}
\end{figure}

Since the Lorentz force is the only possible source of charging outside the core where $|\Delta({\bm r})|={\rm constant}$,
Eq.\ (\ref{Q}) should be quantitatively correct for extreme type-II materials.
Indeed, this contribution can also be understood in terms of Bernoulli's principle $\frac{1}{2}m^*v_{\rm s}^2+e\Phi={\rm constant}$ in the presence of superflow $v_{\rm s}$
with mass $m^*$.\cite{Kita09,AW68}
Note in this context that the constant shift $\Delta\mu\equiv \mu_{\rm s}-\mu_{n}$ 
between the normal and (homogeneous) superconducting states
does not affect ${\bm E}({\bm r})=-{\bm\nabla}\Phi({\bm r})$ and hence does not affect charging at all.
The reduction of $|\Delta({\bm r})|$ for $r\lesssim \xi_0$ may also contribute to the charging when particle-hole asymmetry is present,
as discussed in earlier studies.\cite{KF,Feigel'man}
Since the charge screening length $\lambda_{\rm TF}\sim\hbar/p_{\rm F}$ is short, however, this additional contribution, if any, can only cause extra spatial variation confined 
in $r\lesssim r_{\rm c}$ that cancels out within the core due to the charge neutrality condition.

Choosing $\xi_0 \sim 20 \ ${\AA} and $\lambda_{\rm L}=100\xi_0$ as appropriate values for high-$T_{\rm c}$ superconductors
with the magnetic field along the $c$-axis, we can use Eq.\ (\ref{Q}) to estimate
the vortex-core charge accumulated over the length $\Delta z \sim 5 \ ${\AA} along the flux line as 
$| Q |\equiv | Q_\lambda |\Delta z \sim 10^{-5} |e|$; it is much smaller than the previous estimates $| Q| \sim 10^{-3} |e|$ \cite{KF} and $| Q | \sim 10^{-4} |e|$ \cite{Feigel'man}. 
Note that at the same time, the magnitude can be increased substantially for smaller $\lambda_{\rm L}$ according to Eq.\ (\ref{Q}).

\section{Numerical Examples for Vortex-Core Charging \label{sec:V}}

We have also performed detailed numerical calculations on the following dimensionless
single-particle energy of a two-dimensional square lattice appropriate for high-$T_{\rm c}$ superconductors  \cite{Kontani2,Kita09}:
\begin{align}
\varepsilon_{\bm p}=&\, -2(\cos p_{x}+\cos p_{y})+4t_{1}(\cos p_{x}\cos p_{y}-1)
\nonumber \\
&\,+2t_{2}(\cos 2p_{x}+\cos 2p_{y}-2) 
\label{E_k}
\end{align}
with $t_{1}=1/6$ and $t_{2}=-1/5$, which forms a band  over
$-4\leq \varepsilon_{\bm p}\leq 4$. 
The normal Hall coefficient for this model changes sign from negative to positive as the electron filling $n\in[0,2]$ is increased through $n_{\rm c}=1.03$ \cite{Kita09}.
We study an isolated $d$-wave vortex with ${\bm B}\parallel {\bm z}$ centered at the origin in the $(x,y)$ plane  in the clean limit; 
the pair potential is given by
$\Delta ({\bm p}_{\rm F}, {\bm r}) = \Delta ({\bm r}) \phi ({\bm p}_{\rm F}) e^{- i  \varphi}$, where $\varphi \!\equiv\! \arctan (y / x)$,
and $\phi ({\bm p}_{\rm F}) $ is modeled for $n \gtrsim 0.8$ as $\phi ({\bm p}_{\rm F}) =C \left[ ( p_{{\rm F} x} - \pi )^2 - ( p_{{\rm F} y} - \pi )^2 \right]$ 
 with $C$ denoting the normalization constant determined by $\langle | \phi |^2 \rangle_{\rm F} \!=\! 1$. 
 
Our numerical procedure is summarized as follows. 
We first solve the standard Eilenberger equations self-consistently \cite{Ichioka1,KitaText} to obtain $(g_0,\Delta,{\bm B})$ for the isolated 
$d$-wave vortex.
The resulting solution is used subsequently to calculate the electric field and charge using Eq.\ (\ref{EEq}) and
$\rho=\epsilon_0 {\bm \nabla} \cdot {\bm E}$, respectively.
The parameters of this system
are the coherence length $\xi_0$, magnetic penetration depth $\lambda_{0}\!\equiv\! \left[\mu_0 N(0)e^2\langle v_{\rm F}^2\rangle_{\rm F}\right]^{-1/2}$,
Thomas-Fermi screening length $\lambda_{\rm TF}$, and quasiclassical parameter $\delta$.
We have chosen $\lambda_{0}\!=\!100\xi_0$, $\lambda_{\rm TF}\!=\!0.05\xi_0$, and $\delta\!=\!0.05$ as appropriate values for high-$T_{\rm c}$ superconductors.
The London penetration depth at finite temperatures can be written in terms of $\lambda_0$ by $\lambda_{\rm L} \!=\! \lambda_{0} \langle v_{\rm F} \rangle_{\rm F} \left[2 \langle (1 - Y) v_{{\rm F} x}^2 \rangle_{\rm F} \right]^{-1/2}$.

\begin{figure}[t]
        \begin{center}
                \includegraphics[width=0.95\linewidth]{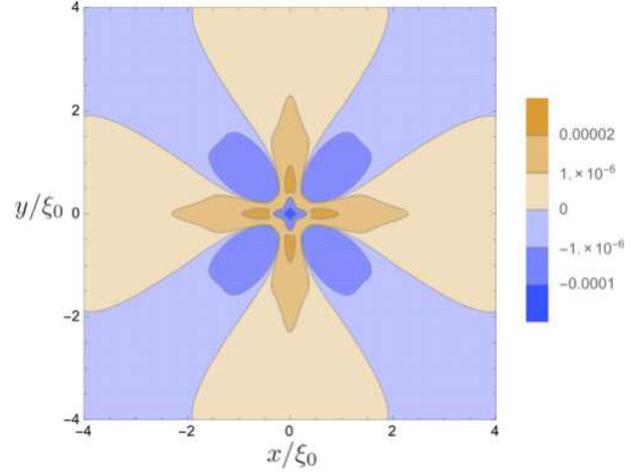}
                \end{center}
\caption{
(Color online) Charge density $\rho ({\bm r})$ in units of $\rho_0 \equiv \epsilon_0 \Delta_0 / | e | \xi_0^2$ over $-4\xi_0 \!\le\! x, y \!\le\! 4\xi_0$ for $n = 0.9$ at $T = 0.2 T_{\rm c}$.
}
\label{fig4}
\end{figure}
\begin{figure}[t]
        \begin{center}
                \includegraphics[width=0.95\linewidth]{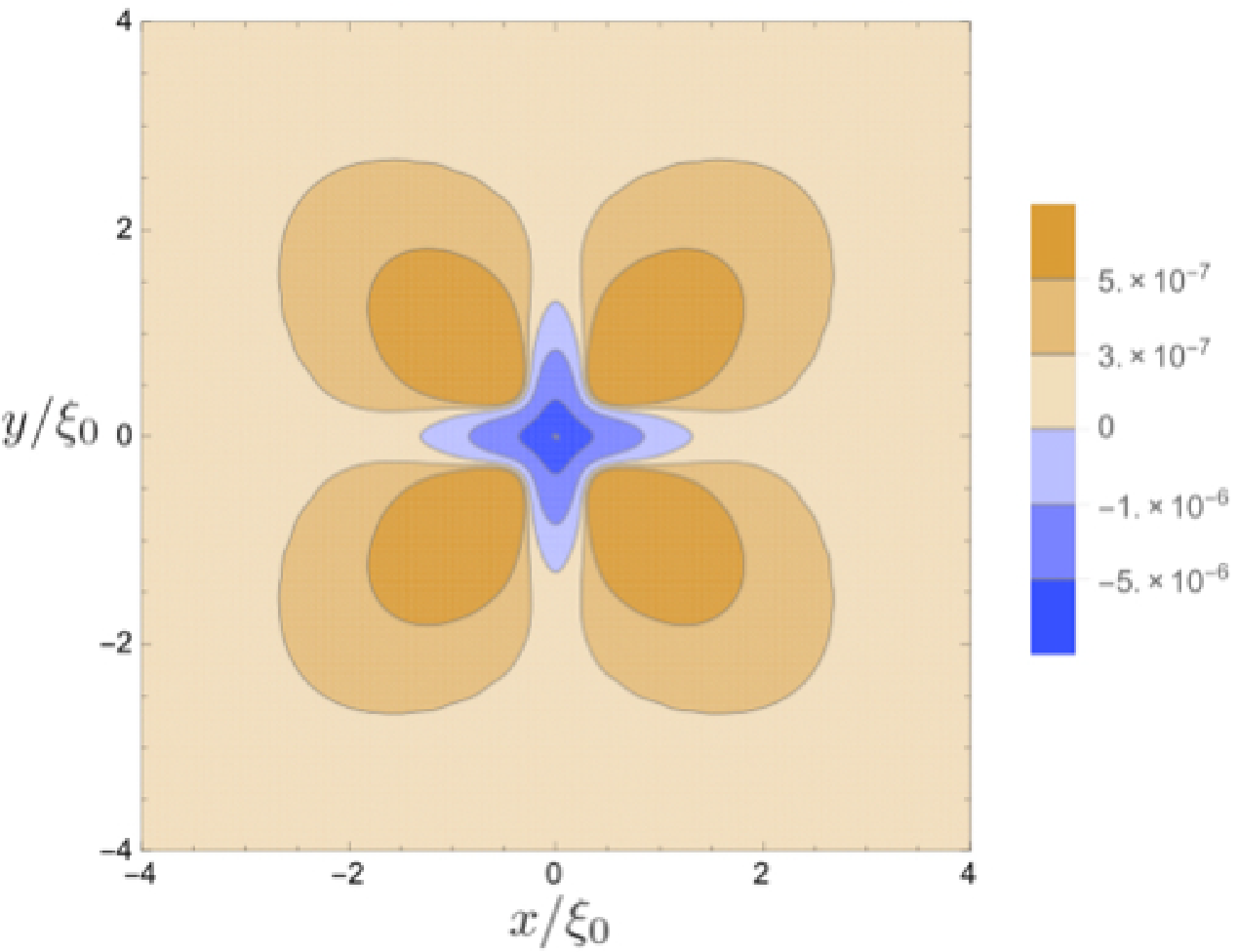}
                \end{center}
\caption{
(Color online) Electric field along the radial direction $E_r ({\bm r})$ in units of $E_0 \equiv \Delta_0 / | e | \xi_0$ over $-4\xi_0 \!\le\! x, y \!\le\! 4\xi_0$ for $n = 0.9$ at $T = 0.2 T_{\rm c}$. }
\label{fig5}
\end{figure}
\begin{figure}[t]
        \begin{center}
                \includegraphics[width=0.95\linewidth]{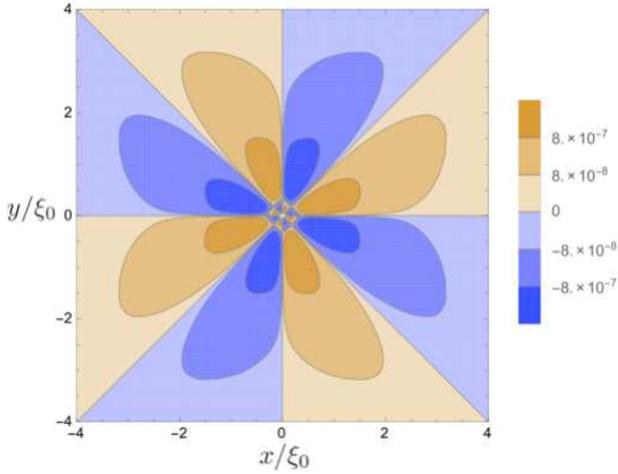}
                \end{center}
\caption{
(Color online) Electric field along the angular direction $E_\varphi ({\bm r})$ in units of $E_0 \equiv \Delta_0 / | e | \xi_0$ over $-4\xi_0 \!\le\! x, y \!\le\! 4\xi_0$ for $n = 0.9$ at $T = 0.2 T_{\rm c}$. }
\label{fig6}
\end{figure}
\begin{figure}[t]
        \begin{center}
                \includegraphics[width=1.0\linewidth]{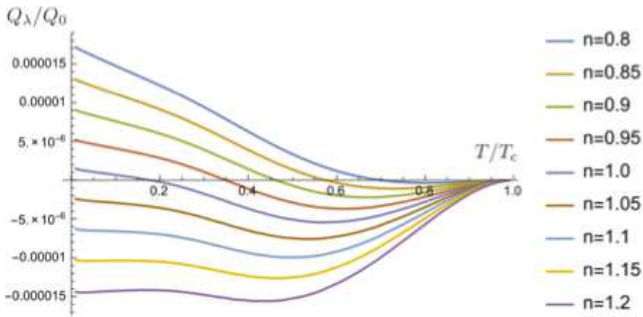}
                \end{center}
\caption{
(Color online) Vortex-core charge $Q_\lambda$ per unit length along the flux line in units of $Q_0 \equiv \epsilon_0 \Delta_0 / | e |$ as a function of temperature over
$0.8 \le n\le 1.2$. }
\label{fig7}
\end{figure}

Figure \ref{fig1} plots the charge density  in the core region for $n=1.95$ with an almost isotropic holelike Fermi surface at $T/T_{\rm c}=0.2$,
where $T_{\rm c}$ denotes the superconducting transition temperature at zero magnetic field.
Here, the fourfold symmetry in the core region is due solely to the gap anisotropy, which becomes obscure outside the core region.
Indeed,  the corresponding distribution for the $s$-wave gap has been confirmed to be completely isotropic. 
The sign of the core charge for this holelike Fermi surface is negative,
as pointed out previously \cite{KF}.
Figures \ref{fig2} and \ref{fig3} plot the electric field of the radial and angular components 
in the core region for $n=1.95$ at $T / T_{\rm c} = 0.2$, respectively. 
The whole sign of the charge density and electric field is reversed for $n=0.05$ with the electron-like Fermi surface.

On the other hand, the charge density for a realistic case of $n=0.9$ exhibits more complicated spatial and temperature dependences.
This filling is close to $n_{\rm c}=1.03$, where the normal Hall coefficient changes its sign, so that
we expect a substantial effect of the Fermi surface anisotropy on the charge distribution according to Eq.\ (\ref{R_H}).
Figure \ref{fig4} plots the charge density in the core region at $T/T_{\rm c}=0.2$. 
Here, the sign of charge at the core center is negative, which is reversed in the adjacent region, and 
the integrated charge over $r \le \xi_0$ and $r \le 2 \xi_0$ is found to be positive. 
Compared with the case of $n=1.95$, the fourfold symmetry is clearer here and extends far outside the core,
which may be attributed to the cooperative effect of the gap and Fermi surface anisotropies.
Figures \ref{fig5} and \ref{fig6} plot the electric field along the radial and angular directions 
in the core region for $n=0.9$ at $T / T_{\rm c} = 0.2$, respectively. 
The sign change of the Hall electric field between the core region and outside the region is caused by the spatial variation in the excitation curvature due to the spatial dependence of the energy gap.

Figure \ref{fig7} plots the temperature dependence of the vortex-core charge $Q_\lambda$ accumulated within $r\leq 2\xi_0$ 
for the fillings $0.8 \!\le\! n\!\le\! 1.2$. 
We observe that both the magnitude and sign of the vortex-core charge change as functions of temperature.
Equation (\ref{R_H}) enables us to attribute this charge to the variation of the excitation curvature under the growing energy gap as $T\rightarrow 0$.
This sign change is beyond the scope of the earlier studies based on the density of states \cite{KF,Feigel'man} and
may be regarded as a definite outcome of our microscopic approach. 
We have confirmed that numerical results can be reproduced quantitatively using Eq.\ (\ref{Q}) with $r_{\rm c}=2\xi_0$.
Finally, note that both the sign and magnitude of the vortex-core charge are detectable by NMR \cite{Kumagai}.

\section{Summary \label{sec:VI}}

We have performed a theoretical study on vortex-core charging.
Our microscopic approach based on the augmented quasiclassical equations has revealed
the essential importance of the Fermi surface curvature and gap anisotropy in determining the sign and magnitude of the vortex-core charge.
We hope that our study will stimulate detailed experiments on vortex-core charging.


\begin{thebibliography}{9}
\bibitem{KF} D. I. Khomskii and A. Freimuth, {Phys. Rev. Lett.} {\bf 75}, 1384 (1995). 
\bibitem{Feigel'man}M. Feigel'man, V. Geshkenbein, A. Larkin, and V. M. Vinokur, JETP Lett. {\bf 62}, 834 (1995).
\bibitem{Blatter} G. Blatter, M. Feigel'man, V. Geshkenbein, A. Larkin, and A. van Otterlo, {Phys. Rev. Lett.} {\bf 77}, 566 (1996). 
\bibitem{ESR}M. Eschrig, J. A. Sauls, and D. Rainer, Phys. Rev. B {\bf 60}, 10447 (1999).
\bibitem{Hayashi} N. Hayashi, M. Ichioka, and K. Machida, {J. Phys. Soc. Jpn.} {\bf 67}, 3368 (1998). 
\bibitem{MH}M. Matsumoto and R. Heeb, Phys. Rev. B {\bf 65}, 014504 (2001). 
\bibitem{Chen} Y. Chen, Z. D. Wang, J. X. Zhu, and C. S. Ting, {Phys. Rev. Lett.} {\bf 89}, 217001 (2002). 
\bibitem{Knapp} D. Knapp, C. Kallin, A. Ghosal, and S. Mansour, Phys. Rev. B {\bf 71}, 064504 (2005).
\bibitem{Kumagai} K. Kumagai, K. Nozaki, and Y. Matsuda, {Phys. Rev. B} {\bf 63}, 144502 (2001). 
\bibitem{GZ} M. Galffy and E. Zirngiebl, {Solid State Commun}, {\bf 68}, 929 (1988). 
\bibitem{Iye} Y. Iye, S. Nakamura, and T. Tamegai, {Physica C} {\bf 159}, 616 (1989). 
\bibitem{AGL} S. N. Artemenko, I. G. Gorlova, and Yu. I. Latyshev, {Phys. Lett. A} {\bf 138}, 428 (1989). 
\bibitem{Hagen1} S. J. Hagen, C. J. Lobb, R. L. Greene, M. G. Forrester, and J. H. Kang, {Phys. Rev. B.} {\bf 41}, 11630 (1990). 
\bibitem{Hagen2} S. J. Hagen, C. J. Lobb, R. L. Greene, and M. Eddy, {Phys. Rev. B.} {\bf 43}, 6246 (1991). 
\bibitem{Chien} T. R. Chien, T. W. Jing, N. P. Ong, and Z. Z. Wang, {Phys. Rev. Lett.} {\bf 66}, 3075 (1991). 
\bibitem{Luo} J. Luo, T. P. Orlando, J. M. Graybeal, X. D. Wu, and R. Muenchausen, {Phys. Rev. Lett.} {\bf 68}, 690 (1992). 
\bibitem{London}F. London, {\it Superfluids} (Dover, New York, 1961), Vol. 1, p. 56.
\bibitem{Kita09}T. Kita, {Phys. Rev. B} {\bf 79}, 024521 (2009).
\bibitem{Eilenberger}G. Eilenberger, Z. Phys. {\bf 214}, 195 (1968).
\bibitem{KP} L. Kramer and W. Pesch, {Z. Phys.} {\bf 269}, 59 (1974). 
\bibitem{Klein} U. Klein, {J. Low Temp. Phys.} {\bf 69}, 1 (1987). 
\bibitem{SM} N. Schopohl and K. Maki, {Phys. Rev. B} {\bf 52}, 490 (1995). 
\bibitem{Ichioka1} M. Ichioka, N. Hayashi, N. Enomoto, and K. Machida, {Phys. Rev. B} {\bf 53}, 15316 (1996). 
\bibitem{SR} J. W. Serene and D. Rainer, Phys. Rep. {\bf 101}, 221 (1983).
\bibitem{LO86} A. I. Larkin and Y. N. Ovchinnikov,
in {\em Nonequilibrium Superconductivity}, ed. D. N. Langenberg and A. I. Larkin (Elsevier, Amsterdam, 1986) Vol. 12, p. 493.
\bibitem{Kopnin} N. B. Kopnin, {\it Theory of Nonequilibrium Superconductivity} (Oxford University Press, New York, 2001). 
\bibitem{KitaText} T. Kita, {\it Statistical Mechanics of Superconductivity} (Springer, Tokyo, 2015).
\bibitem{Kita01} T. Kita, Phys. Rev. B {\bf 64}, 054503 (2001).
\bibitem{AK} E. Arahata and Y. Kato, {J. Low. Temp. Phys.} {\bf 175}, 364 (2014).
\bibitem{AGD} A. A. Abrikosov, L. P. Gor'kov, and I. E. Dzyaloshinski, {\it Methods of Quantum Field Theory in Statistical Physics}
(Prentice Hall, Englewood Cliffs, NJ, 1963). 
\bibitem{Gor'kov} L. P. Gor'kov, Zh. Eksp. Teor. Fiz. {\bf 36}, 1918 (1959) [Sov. Phys. JETP {\bf 9}, 1364 (1959)]; Zh. Eksp. Teor. Fiz. {\bf 37}, 1407 (1959) [Sov. Phys. JETP {\bf 10}, 998 (1960)]. 
\bibitem{Wigner} E. P. Wigner, Phys. Rev. {\bf 40}, 749 (1932). 
\bibitem{Eliashberg} G. M. Eliashberg, Zh. Eksp. Teor. Phys. {\bf 61}, 1254 (1971) [Sov. Phys. JETP {\bf 34}, 668 (1972)].
\bibitem{AW68} C. J. Adkins and J. R. Waldram, Phys. Rev. Lett. {\bf 21}, 76 (1968). 
\bibitem{Kontani2} H. Kontani, Rep. Prog. Phys. {\bf 71}, 026501 (2008).
\end{thebibliography}
\end{document}